# Low-Frequency 1/f Noise Characteristics of Ultra-Thin AlO$_x$-Based Resistive Switching Memory Devices with Magneto-Resistive Responses

**Jhen-Yong Hong** [1,*], **Chun-Yen Chen** [1,2], **Dah-Chin Ling** [1], **Isidoro Martínez** [3,4], **César González-Ruano** [3] and **Farkhad G. Aliev** [3,*]

[1] Department of Physics, Tamkang University, Tamsui Dist., New Taipei City 251301, Taiwan; dcling@mail.tku.edu.tw (D.-C. L.)
[2] Department of Physics, National Central University, Taoyuan City 320317, Taiwan; cychen914@g.ncu.edu.tw (C.-Y.C.)
[3] Depto. Física Materia Condensada, C03, INC and IFIMAC, Universidad Autónoma de Madrid, 28049 Madrid, Spain; isidoro.martinez@uam.es (I.M.); cesar.gonzalez-ruano@uam.es (C.G.-R.)
[4] Faculty of Experimental Sciences, Universidad Francisco de Vitoria, Pozuelo de Alarcón, 28223 Madrid, Spain
\* Correspondence: jyhong@mail.tku.edu.tw (J.-Y.H.); farkhad.aliev@uam.es (F.G.A.)

**Abstract:** Low-frequency 1/f voltage noise has been employed to probe stochastic charge dynamics in AlO$_x$-based non-volatile resistive memory devices exhibiting both resistive switching (RS) and magneto-resistive (MR) effects. A 1/f$^\gamma$ noise power spectral density is observed in a wide range of applied voltage biases. By analyzing the experimental data within the framework of Hooge's empirical relation, we found that the Hooge's parameter $\alpha$ and the exponent $\gamma$ exhibit a distinct variation upon the resistance transition from the low resistance state (LRS) to the high resistance state (HRS), providing strong evidence that the electron trapping/de-trapping process, along with the electric field-driven oxygen vacancy migration in the AlO$_x$ barrier, plays an essential role in the charge transport dynamics of AlO$_x$-based RS memory devices.

**Keywords:** low-frequency 1/f noise; resistive switching; magnetic tunnel junction (MTJ); magneto-resistance (MR); Hooge's parameter

## 1. Introduction

Intensive research interest has been focused on resistive switching (RS) memory devices due to the advantages of non-volatility, high scalability, high integration, and low power consumption, which could be attributed to the simple metal-insulator-metal (MIM) configuration and to their compatibility with the existing complementary metal-oxide-semiconductor (CMOS) technology [1–5]. Despite the fact that RS between the high resistance state (HRS) and low resistance state (LRS) in a variety of insulators, such as perovskite and organics [6,7], realizes the basic function of memory devices, binary metal oxides including TiO$_2$, Al$_2$O$_3$, NiO, and ZnO, among others, have advantages over other complex materials in terms of simplifying the manufacturing process and compatibility of semiconductor processing [1,8,9]. Additionally, binary metal oxides, such aluminum oxides, due to their low-cost fabrication process, self-rectifying properties, and relative wide band gap, have been considered as promising materials for RS memory, exhibiting both bipolar and unipolar RS behaviors [10–12]. Moreover, exploration and manipulation of the transport properties of spin-polarized electrons in Al$_2$O$_3$-based ferromagnet (FM)/insulator/ferromanget (FM) magnetic tunnel junctions (MTJs; identical to RS memories, except for the use of ferromagnetic metals as electrodes) [13,14] offer versatile possibilities for developing novel spintronic applications in information storage and memory evices, which is highly desirable to meet the increasing demand for faster, smaller, and lower energy consumption in the relevant technologies.

To date, numerous studies on MTJs with a thin oxide tunnel barrier have demonstrated that both RS and MR effects can be achieved via electrical control of the resistance states [15–17]. In our recent work [18], devices with ultra-thin (1.5 nm thick) AlO$_x$ barriers showed bipolar resistive switching between HRS and LRS, and simultaneous TMR at HRS and LRS, respectively, making the device a promising candidate for multi-bit memory applications. However, the detailed charge transport dynamics of the AlO$_x$-based RS memory device is still lacking. Note that different types of an alumina barrier, i.e., an oxygen-deficient AlO$_x$ and nearly stoichiometric Al$_2$O$_3$ layers, can be prepared by magnetron sputtering and atomic layer deposition (ALD), respectively. It has been reported that the nearly stoichiometric Al$_2$O$_3$ layer could act as a diffusion barrier for oxygen ions, whereas the oxygen-deficient AlOx layer displays characteristics of charge trapping/de-trapping properties similar to perovskite-type complex oxides [19,20]. The low-frequency noise (LFN) measurement has been widely applied to characterize various disordered systems, which could provide deep insights into the inherent charge transport dynamics related to charge trapping/de-trapping phenomena with randomly distributed trap sites [21,22]. Here, we report an extensive study on the 1/f LFN characteristics of the AlOx-based RS memory devices to address the charge transport dynamics by analyzing the noise power spectrum density (PSD) through Hooge's empirical model [23,24].

## 2. Materials and Methods

The structure of the AlOx-based RS memory device, patterned in a cross-bar configuration with a junction area of 150 μm × 150 μm, was stacked on the glass substrate in the sequence of NiFe (15 nm)/CoFe (10 nm)/AlOx (1.5 nm)/CoFe (35 nm), as shown in the inset of Figure 1a. The bottom NiFe/CoFe layer and top CoFe layer served as the electrode. Since the structure of this device is identical to a MTJ, the bottom NiFe/CoFe also acted as the soft ferromagnetic (FM) electrode and the top CoFe as the hard FM electrode. The entire fabrication process was performed in an ultra-high vacuum (UHV) sputtering chamber with a base pressure of $1 \times 10^{-8}$ mbar. All metallic layers were made with shadow masks by sputtering with an Ar working pressure of $5 \times 10^{-3}$ mbar. The AlOx layer was prepared by nature oxidation of Al, followed by an $O_2$ plasma treatment at 3 mbar for 20 s. A batch of 19 samples were fabricated, which possessed stable sample qualities with similar RS and MR behaviors to our previous work. [18] The current–voltage (I–V) characteristics were measured at various voltages and the magnetoresistance was measured in the current-perpendicular-to-the-plane (CPP) configuration by using the four-probe method with the bias voltage applied on the top electrode while the bottom electrode was grounded. The data acquisition program for the I–V measurements was set to a waiting time of 0.5 s between each voltage step. It requires approximately 2 min to measure an I–V curve.

A schematic circuit diagram for the room-temperature I–V and noise measurements is displayed in Figure 2a. The I–V measurements was taken by using a Keithley 2400 source measure unit (SMU) and a Data Translation DT322 PCI data acquisition (DAQ) board. For the noise measurements, the DC bias was supplied by a Keithley 2400 SMU and the signal associated with the current and noise was amplified by two home-made preamplifiers and then was sent to two commercial SR 560 amplifiers for further amplifications. The noise spectrum was measured with an SR 780 spectrum analyzer. A more detailed description of the circuit design can be found elsewhere [25]. The voltage sweep was designated to stop for the PSD measurement such that the system can relax to an equilibrium state and the desired current/voltage was applied for 10 s before the spectrum analyzer started to take each measurement. It requires about 60 s for one measured PSD (with 200 times of spectrum-averaging). In total, the I–V curve along with the noise measurements requires approximately 40 min. The instrumental set-up was calibrated prior to the noise measurement by measuring the equilibrium Johnson–Nyquist voltage noise, in which $S_V = 4k_BRT$ at room temperature with a series of commercial resistances, where $k_B$ is the Boltzmann constant and R is the sample resistance.

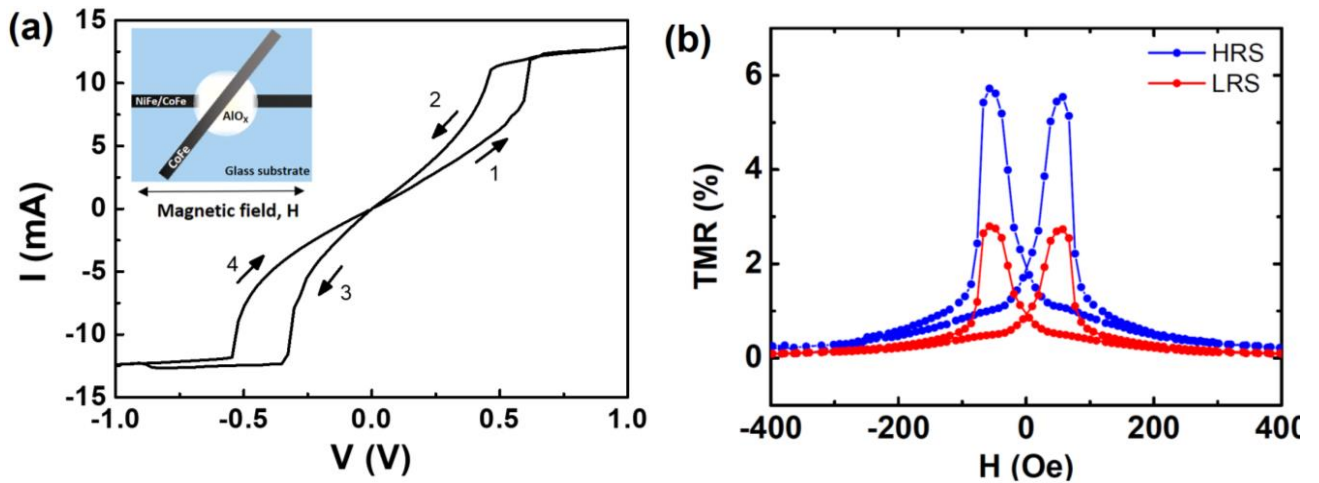

**Figure 1.** (**a**) Current–voltage (I–V) curve of a NiFe/CoFe/AlOx/CoFe RS memory device with arrows indicating the directions of the voltage sweep. Note that the threshold setting and resetting voltage was about 0.6 V and −0.6 V, respectively.

(**b**) The room-temperature TMR as a function of the magnetic field measured at 2 mV for HRS and LRS, respectively. The inset shows the scheme of the device structure.

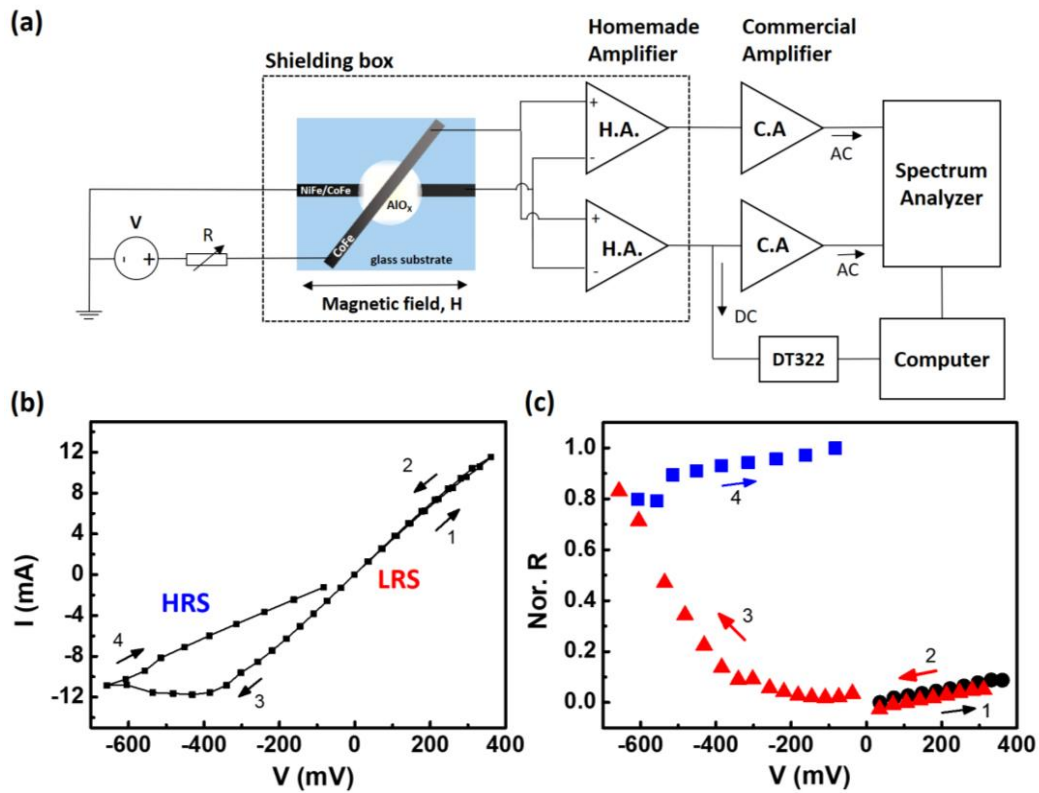

**Figure 2.** (**a**) Experimental setup for LFN measurements. The device and home-made preamplifiers were placed in a shielding box. (**b**) The I–V characteristics of the device with bias voltage in the range of −650 mV to 360 mV. (**c**) Normalized junction resistance of a resistive memory device as a function of the applied bias voltage. In order to better understand the noise characteristics (see Figures 3 and 4), the device was intentionally controlled in the LRS by applying a positive voltage lower than the threshold setting voltage of ~0.6 V. The arrow shows the direction of the voltage sweep.

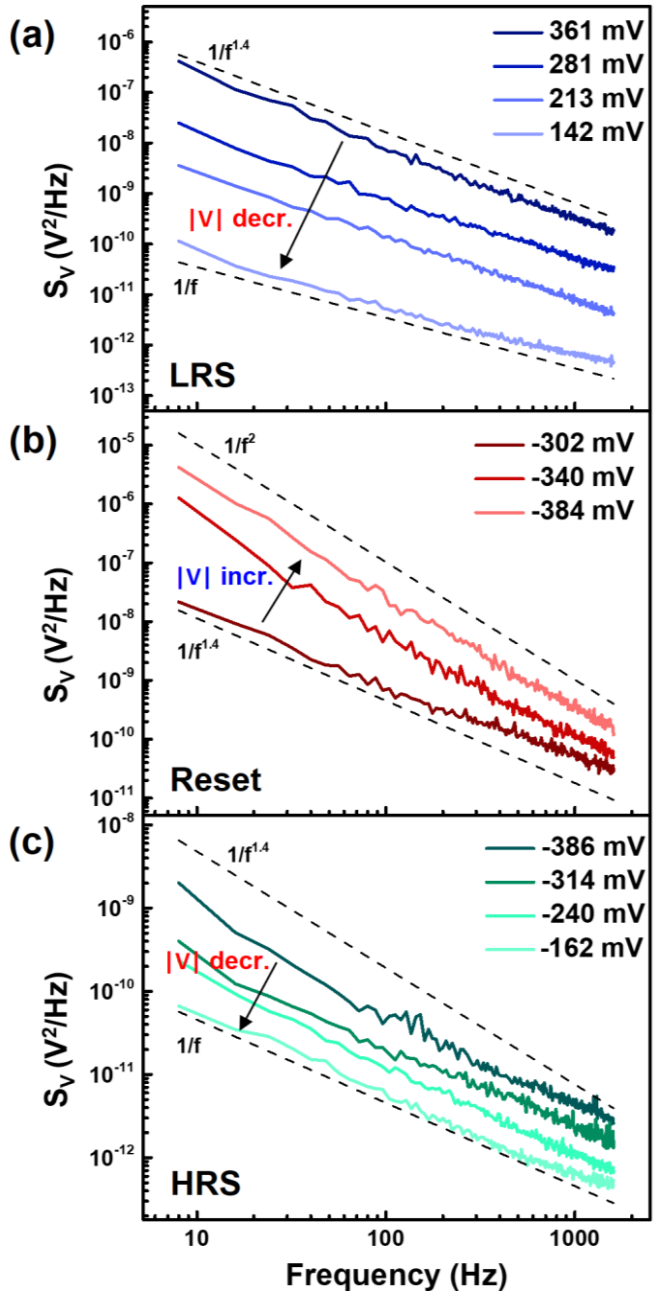

**Figure 3.** The PSD of the devices measured at the (**a**) LRS in step 2, (**b**) reset process in step 3, and (**c**) HRS in step 4. As the device underwent a transition from LRS to HRS in step 3, the exponent of 1/f increased dramatically from 1.4 to 2.0, which shows a significant discrepancy as compared to (**a**) and (**b**).

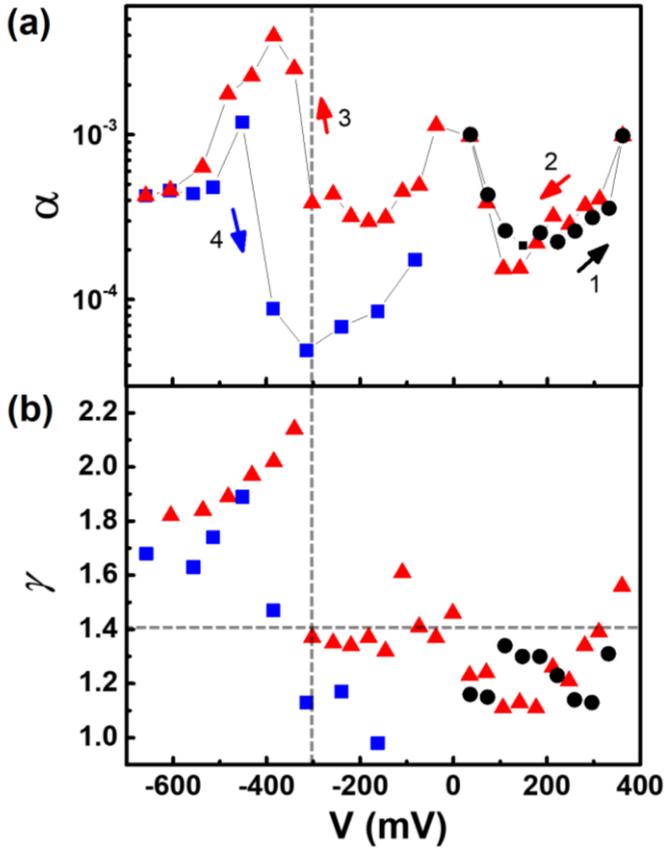

**Figure 4.** The bias voltage dependence of (**a**) Hooge's parameter $\alpha$ and (**b**) the exponent $\gamma$ extracted from noise PSD based on Hooge's empirical relation.

## 3. Results and Discussion

The I–V curve of the device investigated is shown in Figure 1 and exhibited reversible, non-volatile bipolar resistive switching characteristics. After the electro-forming process, the voltage was swept between +1.0 and −1.0 V in a cyclic manner, with the arrows indicating the direction of the voltage sweep. When the voltage increased gradually from 0 to +1.0 V (step 1), the current increased abruptly at ~0.6 V, which transited the device to LRS, followed by a constant current trace limited by setting the compliance current of 13 mA. On the negative bias side (step 3), a current jump was observed at ~−0.32 V, which switched the device from LRS to HRS. Here, we designated LRS as the "ON" or "1" state and HRS as the "OFF" or "0" state. The I–V curve clearly shows the RS memory characteristics. Low-switching voltage characteristics with 7.32 mW of set power (12 mA @ 0.61 V) and 4.20 mW of reset power (−12 mA @ -0.35 V) were obtained in this device. Note that the value of the switching voltage was the same order of magnitude as that of devices with a large junction area. As compared to our previous study [18], the low set and reset voltages in this device could be attributed to the material and interfacial properties associated with the higher level of the compliance current, which in turn shifts the switching threshold voltage [26]. Previous studies have reported that a smooth switching of the I–V curve could be interpreted as a modification of the Schottky barrier height at the metal–oxide interface, which is responsible for the trapping/de-trapping of electrons in the interfacial-type RS memory devices [27,28].

Figure 1b shows the magneto-transport characterization of the AlO$_x$-based RS device, in which the cycling of the junction resistance is plotted with the applied in-plane magnetic field. The resistance curves display the typical MTJ characteristics at room temperature with a tunneling magneto-resistive (TMR) ratio defined as

$$\text{TMR\%} = \frac{\Delta R}{R_P} \times 100\% = \frac{R_{AP} - R_P}{R_P} \times 100\% \quad (1)$$

where $R_P$ and $R_{AP}$ depict the resistances when the magnetizations of the FM electrodes are in the parallel and anti-parallel configurations, respectively. To better understand the TMR ratio associated with the HRS and LRS as a function of the magnetic field in the AlO$_x$ based MTJ, the device was set from one resistive state to another by applying voltages of 1.0 V (set) and −1.0 V (reset), and the magnetoresistance measurements were performed subsequently at 2 mV. Figure 1b shows the MR loop of the MTJ at both HRS and LRS with a TMR ratio of 5.8% at HRS and 2.7% at LRS, indicating that both the RS and MR effects were present in a single device, manifesting as four electrically and magnetically tunable resistance states. The consistent behavior was observed in a number of MTJs made under identical conditions, albeit with some variations in the resistance value. Compared to our previous study [18], the setting of a higher compliance current in this work reduced the junction resistance and thereby lowered the tunneling probability, which resulted in a lower TMR ratio in both HRS and LRS. Notably, the ability to control the RS and MR performance is desirable for future applications of the electrical control of magnetism and emerging spintronic devices, such as spin-orbit-toque-induced magnetization switching [29,30].

As for the noise characteristics, the same device was carried to the LFN measurement system to measure the noise PSD as a function of the applied voltage. The device was intentionally kept at LRS (steps 1 and 2) in the initial state, as shown in Figure 2b. As the voltage increased from 0 to +360 mV (step 1) and then swept back to 0 V (step 2), the hysteresis-free I–V curve was observed, as shown in Figure 2b. This indicates that the device remained in the LRS and the Joule heating effect was negligible when the voltage was swept between 0 and 360 mV. Further investigations showed that as long as the applied voltage did not go beyond the threshold set voltage of 0.6 V, the device consistently remains in the LRS. However, when the negative bias voltage was applied, the resistance of the device gradually increased up to V ~−300 mV and then underwent a transition to the HRS at V ~−600 mV, indicated as the reset process (step 3) in Figure 2b. As the voltage was swept back to zero, the device then remained in the HRS. Figure 2c presents the normalized resistance as a function of the applied voltage, which evidently shows the "ON" (LRS, step 1) and "OFF" (HRS, step 4) states upon sweeping the voltage.

Figure 3 displays the representative PSD recorded in the frequency range of 8–1600 Hz for different resistance states during the voltage sweep in Figure 2b. The 1/f-like low-frequency noise PSD can be characterized by Hooge's empirical expression [23,24]

$$S_V = \alpha \frac{V^2}{A \cdot f^\gamma} \qquad (2)$$

where $A$ is the cross-sectional area of the junction, $V$ is the voltage across the junction, $\gamma$ is the exponent of the 1/f noise spectrum, and $\alpha$ is the Hooge's parameter, which is a measure of noise amplitude. Both the Hooge's parameter $\alpha$ and the 1/f exponent $\gamma$ could be extracted from the linear fitting to log(S)-versus-log(f) curves. Figure 3a shows the PSD recorded in the LRS (step 2) at various bias voltages. It was seen that for the device in the LRS, the exponent index $\gamma$ was between 1.4 and 1.0 as the voltage ranged from 361 mV to 142 mV. Similar LFN features with the exponent $\gamma$ changing with the applied voltage have been observed in other RS memory devices [31,32]. As the voltage further swept to the negative region (step 3) across the reset process ~−384 mV, the exponent $\gamma$ exhibited a significant increase up to 2.0, as illustrated in Figure 3b. The higher exponent value in the reset process could have been caused by the Brownian motion or the diffusion of oxygen ions during the strong trapping/de-trapping processes, which induce charge number fluctuations [33] . Moreover, the device was set to the HRS and the magnitude of the applied voltage decreased from −386 mV to a lower level of −162 mV (step 4), while the value of exponent $\gamma$ progressively reduced from 1.4 to 1.0 and therefore showed a typical 1/f dependence in the PSD (see Figure 3c).

To obtain a clearer picture of the LFN characteristics, the trajectory of the Hooge's parameter $\alpha$ and the exponent $\gamma$, obtained from fitting Equation (2) to the experimental data of the 1/f noise spectrum measured at different bias voltages, is presented in Figure 4. The arrows indicate the voltage sweeping direction. Figure 4a outlines the bias voltage dependence of the Hooge's parameter of the device, which qualitatively elucidates how noise amplitude varies with bias voltage when the device was in different resistance states. The $\alpha$ remains approximately at the same value for a fixed bias voltage in steps 1 and 2, which in turn demonstrates that the device was well kept at the LRS, in good agreement with the hysteresis-free I–V curve with a voltage sweep between 150 mV and 360 mV, as displayed in Figure 2b. Furthermore, as the voltage further decreased to the lower bias region (~150 mV to −250 mV), the value of $\alpha$ increased with lowering bias, which is a typical behavior observed in the MTJs [34]. While the device underwent the reset process (step 3) at ≅ −300 mV, the $\alpha$ increased dramatically. Followed by the reset process, the value of $\alpha$ decreased and remained at the same value up to

−600 mV. By comparing the extracted Hooge's parameter $\alpha$ in LRS and HRS, we can see that $\alpha$ was one order of magnitude higher in LRS than that in HRS. In the reset region, unlike the LRS and HRS, $\alpha$ was relatively high along with the observed $1/f^{2.0}$ noise characteristics, suggesting that the main noise source in the reset region was different from that of the HRS or LRS, which could be attributed to the Brownian motion or the diffusion of oxygen ions in a process dominated by Joule heating as the device undergoes the resistance state transition [33].

Figure 4b summarizes the exponent $\gamma$ of the $1/f^\gamma$-like noise spectrum for the device, making a transition from LRS (step 1 and 2) to reset (step 3) and then to HRS (step 4). The PSD of the device in the LRS exhibited an exponent of $1.0 \leq \gamma \leq 1.4$ until the device was biased through the reset process at V $\cong$ −300 mV (in step 3), where the exponent $\gamma$ revealed a noticeable increase. The increase of the exponent from the LRS to the reset process could be attributed to the onset of the trapping/de-trapping of charges in deep trap levels [31]. When bias voltage was slightly less than −300 mV, the exponent $\gamma$ gradually decreased and remained in the range of 1.8 to 2.2 when the bias was less than −350 mV. When the device was switched to HRS (step 4) at V $\geq$ −600 mV, the exponent was in the range of 1.5 to 1.9, possibly arising from the non-vanishing trapping/de-trapping of charges. As the applied voltage further increased to a value higher than −400 mV, the exponent gradually decreased to the range of 1.0 to 1.4, exhibiting normal 1/f characteristics.

Previous studies have shown that the sputtered AlOx barrier contains inevitable oxygen deficiencies, giving rise to intrinsic oxygen-vacancy-rich amorphous structures [20,35–37]. In our device, the active layer with the ultra-thin AlOx barrier could be modeled as a quasi-two-dimensional system, in which an artificial gradient of the oxygen vacancy concentration is established during the process of the plasma oxidation of the Al layer. As a result, the concentration of the oxygen vacancies in the AlOx layer would be higher near the bottom electrode, which could contribute to the smooth switching I–V characteristics of an interfacial-type RS memory, as illustrated in Figure 1a. By applying an electric field, the oxygen vacancy migration and the trapping/de-trapping process of the injected electrons would play an important role in the resistive switching behavior of devices, which is consistent with previous reports [20,35–38]. Based on this scenario, it is more probable that in the LRS, the vacancies were densely distributed near the conductive path formed by oxygen vacancies inside the AlOx barrier and therefore the PSD exhibited 1/f-like dependence. However, an enhancement of the Hooge's parameter $\alpha$ and exponent $\gamma$ in the reset process could be attributed to the strong trapping/de-trapping of the injected electrons, along with the electric field-driven oxygen vacancy migration [32,33]. Upon the reset process, there were more dominant vacancies near the conductive path, resulting in the PSD being inversely proportional to the square of the frequency. As the device was switched to HRS, the number of vacancies near the conductive path reduced and distributed more uniformly, leading to an envelope of 1/f dependence in the frequency domain in the lower bias region.

## 4. Conclusions

In summary, we have investigated LFN characteristics in AlOx-based memory devices with both RS and MR effects. By analyzing the noise PSD within the framework of Hooge's empirical relation, we find that the Hooge's parameter $\alpha$ and the exponent $\gamma$ exhibit a distinct variation upon resistance transition from the LRS to the HRS, providing strong evidence that the electron trapping/de-trapping process, along with the electric field-driven oxygen vacancy migration in the AlOx barrier, plays an essential role in the charge-transport dynamics of AlOx-based RS memory devices.

**Author Contributions:** Conceptualization, J.-Y.H. and F.G.A.; data acquisition, C.-Y.C., I.M., and C.G.-R.; methodology, J.-Y.H., D.-C.L., and F.G.A.; writing—original draft preparation, J.-Y.H.; writing—review and editing, J.-Y.H., D.-C.L., and F.G.A. All authors have read and agreed to the published version of the manuscript.

**Funding:** This research was funded by Ministry of Science and Technology (MOST) of Taiwan under grant number 106-2112-M-032-001-MY3 and Spanish Ministerio de Ciencia (RTI2018-095303-B-C55 and CEX2018-000805-M) and Comunidad de Madrid (NANOMAGCOST-CM, P2018/NMT-4321) grants.

**Acknowledgments:** JYH acknowledges Minn-Tsong Lin from the Department of Physics at National Taiwan University for the support during the sample preparation and Ping-Hung Yeh from the Department of Physics at Tamkang University for the support when taking optical microscope (OM) and scanning electron microscope (SEM) images. FGA acknowledges "Acción financiada por la Comunidad de Madrid en el marco del convenio plurianual con la Universidad Autónoma de Madrid en Línea 3: Excelencia para el Profesorado Universitario".

**Conflicts of Interest:** The authors declare no conflict of interest.

## References

1. Slesazeck, S.; Mikolajick, T. Nanoscale resistive switching memory devices: A review. *Nanotechnology* **2019**, *30*, 352003.
2. Waser, R.; Aono, M. Nanoscience and Technology: A Collection of Reviews from Nature Journals. *World Sci*. **2010**, 158–165.
3. Kwon, D.H.; Kim, K.M.; Jang, J.H.; Jeon, J.M.; Lee, M.H.; Kim, G.H.; Li, X.S.; Park, G.S.; Lee, B.; Han, S.; et al. Atomic structure of conducting nanofilaments in $TiO_2$ resistive switching memory. *Nat. Nanotechnol*. **2010**, *5*, 148.
4. Ielmini, D. Resistive switching memories based on metal oxides: mechanisms, reliability and scaling. *Semicond. Sci. Technol*. **2016**, *31*, 063002.
5. Yang, J.J.; Strukov, D.B.; Stewart, D.R. Memristive devices for computing. *Nat. Nanotechnol*. **2013**, *8*, 13–24.
6. Gao, S.; Yi, X.; Shang, J.; Liu, G.; Li, R.W. Organic and hybrid resistive switching materials and devices. *Chem. Soc. Rev.* **2019** *48*, 1531–1565.
7. Panda, D.; Tseng, T.Y. Perovskite oxides as resistive switching memories: A review. *Ferroelectrics* **2014**, *471*, 23–64.
8. Sawa, A. Resistive switching in transition metal oxides. *Mater. Today* **2008**, *11*, 28–36.
9. Lee, J.S.; Lee, S.; Noh, T.W. Resistive switching phenomena: A review of statistical physics approaches. *Appl. Phys. Rev.* **2015**, *2*, 031303.
10. Sarkar, B.; Lee, B.; Misra, V. Understanding the gradual reset in $Pt/Al_2O_3/Ni$ RRAM for synaptic applications. *Semicond. Sci. Technol*. **2015**, *30*, 105014.
11. Jang, J.; Choi, H.H.; Paik, S.H.; Kim, J.K.; Chung, S.; Park, J.H. Highly improved switching properties in flexible aluminum oxide resistive memories based on a multilayer device structure. *Adv. Electron. Mater*. **2018**, *4*, 1800355.
12. Tran, X.A.; Zhu, W.; Liu, W.J.; Yeo, Y.C.; Nguyen, B.Y.; Yu, H.Y. A Self-Rectifying $AlO_y$ Bipolar RRAM with Sub-50-μA Set/Reset current for cross-bar architecture. *IEEE Electron Device Lett*. **2012**, *33*, 1402–1404.
13. Miyazaki, T.; Tezuka, N. Spin polarized tunneling in ferromagnet/insulator/ferromagnet junctions. *J. Magn. Magn. Mater*. **1995** *151*, 403–410.
14. Moodera, J.S.; Kinder, L.R. Ferromagnetic–insulator–ferromagnetic tunneling: Spin-dependent tunneling and large magnetoresistance in trilayer junctions. *J. Appl. Phys*. **1996**, *79*, 4724–4729.
15. Teixeira, J.M.; Ventura, J.; Fermento, R.; Araujo, J.P.; Sousa, J.B.; Wisniowski, P.; Freitas, P.P. Electroforming, magnetic and resistive switching in MgO-based tunnel junctions. *J. Phys. D Appl. Phys*. **2009**, *42*, 105407.
16. Yoshida, C.; Kurasawa, M.; Lee, Y.M.; Aoki, M.; Sugiyama, Y. Unipolar resistive switching in CoFeB/ MgO/ CoFeB magnetic tunnel junction. *Appl. Phys. Lett*. **2008**, *92*, 113508.
17. Krzysteczko, P.; Reiss, G.; Thomas, A. Memristive switching of MgO based magnetic tunnel junctions. *Appl. Phys. Lett*. **2009**, *95*, 112508.
18. Hong, J.Y.; Hung, C.F.; Yang, K.H.O.; Chiu, K.C.; Ling, D.C.; Chiang, W.C.; Lin, M.T. Electrically programmable magnetoresistance in $AlO_x$-based magnetic tunnel junctions. *Sci. Rep*. **2021**, *11*, 1–7.
19. Persson, K.M.; Ram, M.S.; Wernersson, L.E. Ultra-Scaled $AlO_x$ Diffusion Barriers for Multibit $HfO_x$ RRAM Operation. *IEEE J. Electron Devices Soc*. **2021**, 9, 564–569.
20. Cho, S.; Jung, J.; Kim, S.; Pak, J.J. Conduction mechanism and synaptic behaviour of interfacial switching $AlO_δ$-based RRAM. *Semicond. Sci. Technol*. **2020**, *35*, 085006.
21. Yu, S.; Jeyasingh, R.; Wu, Y.; Wong, H.S.P. Characterization of low-frequency noise in the resistive switching of transition metal oxide $HfO_2$. *Phys. Rev. B* **2012**, *85*, 045324.
22. Kim, Y.; Song, H.; Kim, D.; Lee, T.; Jeong, H. Noise characteristics of charge tunneling via localized states in metal-molecule- metal Junctions. *ACS Nano* **2010**, *4*, 4426–4430.
23. Hooge, F.N.; Kleinpenning, T.G.M.; Vandamme, L.K.J. Experimental studies on 1/f noise. *Rep. Prog. Phys*. **1981**, *44*, 479.
24. Hooge, F.N.; Kleinpenning, T.G.M.; Vandamme, L.K.J. 1/f Noise Sources. *IEEE Trans. Electron. Devices* **1994**, *41*, 1926–1935.
25. Aliev, F.; Cascales, J.P. *Experimental Methods. Noise in Spintronics: From Understanding to Manipulation*; CRC Press: Boca Raton, FL, USA, 2018.
26. Lentz, F.; Roesgen, B.; Rana, V.; Wouters, D.J.; Waser, R. Current Compliance-Dependent Nonlinearity in $TiO_2$ ReRAM. *IEEE Electron. Device Lett*. **2013**, *34*, 996–998.
27. Sassine, G.; La Barbera, S.; Najjari, N.; Minvielle, M.; Dubourdieu, C.; Alibart, F. Interfacial versus filamentary resistive switching in $TiO_2$ and $HfO_2$ devices. *J. Vac. Sci. Technol. B Nanotechnol. Microelectron. Mater. Process. Meas. Phenom*. **2016**, *34*, 012202.
28. Bagdzevicius, S.; Maas, K.; Boudard, M.; Burriel, M. Interface-type resistive switching in perovskite materials. *J. Electroceramics* **2017**, *39*, 157–184.
29. Shaibo, J.; Yang, R.; Wang, Z.; Huang, H.M.; Xiong, J.; Guo, X. Electric field control of resistive switching and magnetization in epitaxial $LaBaCo_2O_{5+δ}$ thin films. *Phys. Chem. Chem. Phys*. **2019**, *21*, 8843–8848.
30. Huang, Q.; Dong, Y.; Zhao, X.; Wang, J.; Chen, Y.; Bai, L.; Dai, Y.; Dai, Y.; Yan, S.; Tian, Y. Electrical Control of Perpendicular Magnetic Anisotropy and Spin-Orbit Torque-Induced Magnetization Switching. *Adv. Electron. Mater*. **2020**, *6*, 1900782.

31. Song, Y.; Jeong, H.; Jang, J.; Kim, T.Y.; Yoo, D.; Kim, Y.; Jeong, H.; Lee, T. 1/f Noise scaling analysis in unipolar-type organic nanocomposite resistive memory. *ACS Nano* **2015**, *9*, 7697–7703.
32. Song, Y.; Lee, T. Electronic noise analyses on organic electronic devices. *J. Mater. Chem. C* **2017**, *5*, 7123–7141.
33. Rocha, P.R.; Gomes, H.L.; Vandamme, L.K.; Chen, Q.; Kiazadeh, A.; De Leeuw, D.M.; Meskers, S.C. Low-frequency diffusion noise in resistive-switching memories based on metal–oxide polymer structure. *IEEE Trans. Electron. Devices* **2012**, *59*, 2483–2487.
34. Almeida, J.M.; Wisniowski, P.; Freitas, P.P. Low-frequency noise in MgO magnetic tunnel junctions: Hooge's parameter dependence on bias voltage. *IEEE Trans. Magn.* **2008**, *44*, 2569–2572.
35. Huang, X.D.; Li, Y.; Li, H.Y.; Xue, K.H.; Wang, X.; Miao, X.S. Forming-free, fast, uniform, and high endurance resistive switching from cryogenic to high temperatures in W/AlO$_x$/Al$_2$O$_3$/Pt bilayer memristor. *IEEE Electron. Device Lett.* **2020**, *41*, 549–552.
36. Wang, Z.; Sun, B.; Ye, H.; Liu, Z.; Liao, G.; Shi, T. Annealed AlO$_x$ film with enhanced performance for bipolar resistive switching memory. *Appl. Surf. Sci.* **2021**, *546*, 149094.
37. Momida, H.; Nigo, S.; Kido, G.; Ohno, T. Effect of vacancy-type oxygen deficiency on electronic structure in amorphous alumina. *Appl. Phys. Lett.* **2011**, *98*, 042102.
38. Kubota, M.; Nigo, S.; Kato, S.; Amemiya, K. Conduction band caused by oxygen vacancies in aluminum oxide for resistance random access memory. *J. Appl. Phys.* **2012**, *112*, 033711.